\documentclass[sigconf]{acmart}

\usepackage{booktabs}
\usepackage{algorithm,algorithmic}
\usepackage{dsfont}
\usepackage{multirow}
\usepackage[english]{babel}
\usepackage{pgfplots}
\usepackage[utf8]{inputenc}
\usepackage{lipsum}
\usepackage{tikz}
\graphicspath{ {images/} }

\def\ind{\mathds{1}}
\def\E{\mathbb{E}}
\def\P{\mathbb{P}}
\def\Cov{\mathrm{Cov}}
\def\Var{\mathrm{Var}}
\def\deg{\mathrm{deg}}
\def\cV{\mathcal{V}}
\def\cN{\mathcal{N}}
\def\co{\mathrm{co}}

\newcommand\independent{\protect\mathpalette{\protect\independenT}{\perp}}
\def\independenT#1#2{\mathrel{\rlap{$#1#2$}\mkern2mu{#1#2}}}

\def\myplotwidth{0.23\textwidth}

\copyrightyear{2019} 
\acmYear{2019} 
\setcopyright{acmlicensed}
\acmConference[KDD '19]{The 25th ACM SIGKDD Conference on Knowledge Discovery and Data Mining}{August 4--8, 2019}{Anchorage, AK, USA}
\acmBooktitle{The 25th ACM SIGKDD Conference on Knowledge Discovery and Data Mining (KDD '19), August 4--8, 2019, Anchorage, AK, USA}
\acmPrice{15.00}
\acmDOI{10.1145/3292500.3330995}
\acmISBN{978-1-4503-6201-6/19/08}

\begin{document}
\title{Estimating Graphlet Statistics via Lifting}

\author{Kirill Paramonov}
\affiliation{
    \institution{Google}
    \city{San Bruno}
    \state{CA}
    \country{USA}
    }
\email{kir.paramonov@gmail.com}

\author{Dmitry Shemetov}
\affiliation{
    \institution{University of California, Davis}
    \city{Davis}
    \state{CA}
    \country{USA}
    }
\email{dshemetov@ucdavis.edu}

\author{James Sharpnack}
\affiliation{
    \institution{University of California, Davis}
    \city{Davis}
    \state{CA}
    \country{USA}
    }
\email{jsharpna@gmail.com}

\begin{abstract}
    Exploratory analysis over network data is often limited by the ability to efficiently calculate graph statistics, which can provide a model-free understanding of the macroscopic properties of a network.
    We introduce a framework for estimating the graphlet count---the number of occurrences of a small subgraph motif (e.g. a wedge or a triangle) in the network.
	For massive graphs, where accessing the whole graph is not possible, the only viable algorithms are those that make a limited number of vertex neighborhood queries.
    We introduce a Monte Carlo sampling technique for graphlet counts, called {\em Lifting}, which can simultaneously sample all graphlets of size up to $k$ vertices for arbitrary $k$.
    This is the first graphlet sampling method that can provably sample every graphlet with positive probability and can sample graphlets of arbitrary size $k$.
    We outline variants of lifted graphlet counts, including the ordered, unordered, and shotgun estimators, random walk starts, and parallel vertex starts. 
	We prove that our graphlet count updates are unbiased for the true graphlet count and have a controlled variance for all graphlets.
	We compare the experimental performance of lifted graphlet counts to the state-of-the art graphlet sampling procedures: Waddling and the pairwise subgraph random walk.


\end{abstract}

%
%


\maketitle

	\section{Introduction}
	\label{sec:intro}
	
    In 1970, \cite{davis1970clustering} discovered that transitivity---the tendency of friends of friends to be friends themselves---is a prevalent feature in social networks.
    Since that early discovery, real-world networks have been observed to have many other common macroscopic features, and these discoveries have led to probabilistic models for networks that display these phenomena.
    The observation that transitivity and other common subgraphs are prevalent in networks motivated the exponential random graph model (ERGM) \cite{frank1986markov}.
    \cite{barabasi1999emergence} demonstrated that many large networks display a scale-free power law degree distribution, and provided a model for constructing such graphs.
    Similarly, the small world phenomenon---that networks display surprisingly few degrees of separation---motivated the network model in \cite{watts1998collective}.
    While network science is often driven by the observation and modelling of common properties in networks, it is incumbent on the practicing data scientist to explore network data using statistical methods. 
    
    One approach to understanding network data is to fit free parameters in these network models to the data through likelihood-based or Bayesian methods \cite{wasserman1996logit,snijders2002markov}. 
	Network statistics, such as the clustering coefficient, algebraic connectivity, and degree sequence, are more flexible tools.
	A good statistic can be used to fit and test models, for example, \cite{watts1998collective} used the local clustering coefficient, a measure of the number of triangles relative to wedges, to test if a network is a small-world graph.
	It was discovered that re-occurring subgraph patterns can be used to differentiate real-world networks, and that genetic networks, neural networks, and internet networks all presented different common interconnectivity patterns, \cite{milo2002network}.
	In this work, we will propose a new method for counting the occurrences of any subgraph pattern, otherwise known as {\em graphlets}---a term coined in \cite{prvzulj2004modeling}---or motifs.
	
\begin{figure}[th]
\begin{center}
\hspace*{\fill}
\minipage{0.06\textwidth}
  \includegraphics[width=\linewidth]{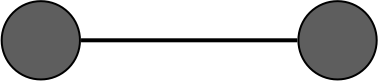}\vspace{-.4cm}
  \caption*{$H_1^{(2)}$}
\endminipage\hfill
\minipage{0.06\textwidth}
  \includegraphics[width=\linewidth]{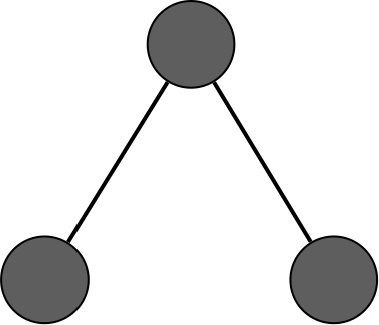}\vspace{-.4cm}
  \caption*{$H_1^{(3)}$}
\endminipage\hfill
\minipage{0.06\textwidth}
  \includegraphics[width=\linewidth]{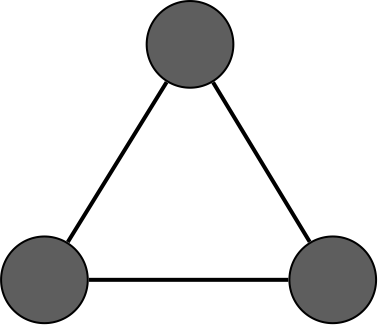}\vspace{-.4cm}
  \caption*{$H_2^{(3)}$}
\endminipage
\hspace*{\fill}
\vskip 5pt
\hspace*{\fill}
\minipage{0.06\textwidth}%
  \includegraphics[width=\linewidth]{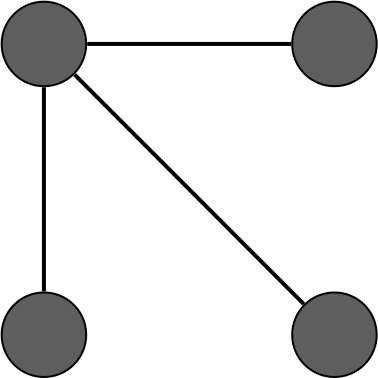}\vspace{-.3cm}
  \caption*{$H_1^{(4)}$}
\endminipage\hfill
\minipage{0.06\textwidth}
  \includegraphics[width=\linewidth]{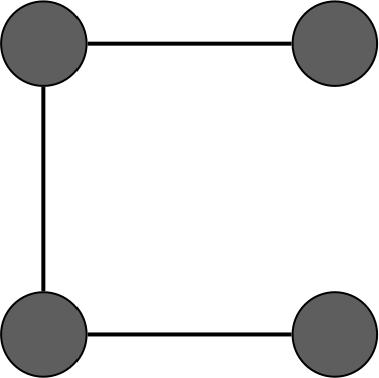}\vspace{-.3cm}
  \caption*{$H_2^{(4)}$}
\endminipage\hfill
\minipage{0.06\textwidth}
  \includegraphics[width=\linewidth]{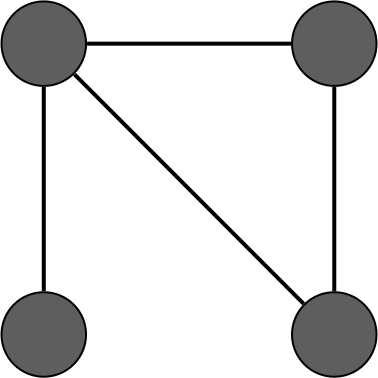}\vspace{-.3cm}
  \caption*{$H_3^{(4)}$}
\endminipage\hfill
\minipage{0.06\textwidth}
  \includegraphics[width=\linewidth]{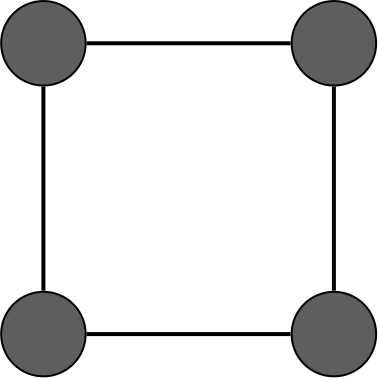}\vspace{-.3cm}
  \caption*{$H_4^{(4)}$}
\endminipage\hfill
\minipage{0.06\textwidth}%
  \includegraphics[width=\linewidth]{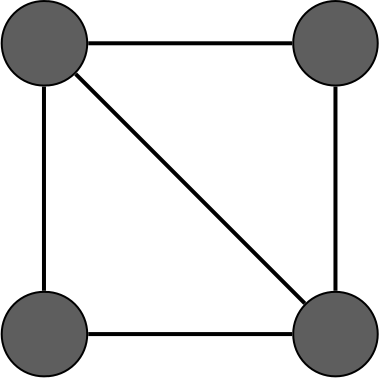}\vspace{-.3cm}
  \caption*{$H_5^{(4)}$}
\endminipage\hspace*{\fill}
\minipage{0.06\textwidth}%
  \includegraphics[width=\linewidth]{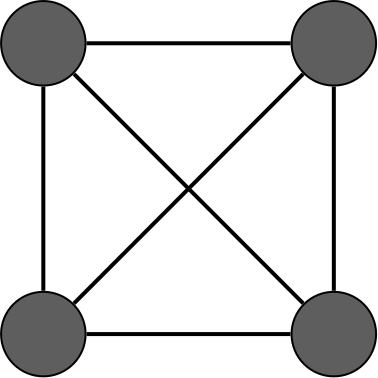}\vspace{-.3cm}
  \caption*{$H_6^{(4)}$}
\endminipage
\hspace*{\fill}
\vskip 5pt
\hspace*{\fill}
\minipage{0.08\textwidth}%
  \includegraphics[width=\linewidth]{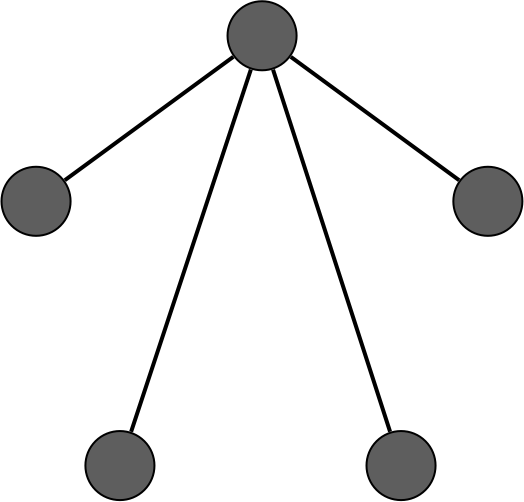}\vspace{-.3cm}
  \caption*{$H_1^{(5)}$}
\endminipage\hfill
\minipage{0.08\textwidth}
  \includegraphics[width=\linewidth]{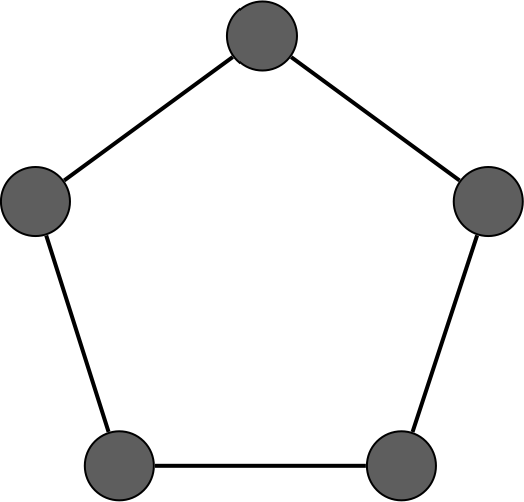}\vspace{-.3cm}
  \caption*{$H_8^{(5)}$}
\endminipage\hfill
\minipage{0.08\textwidth}
  \includegraphics[width=\linewidth]{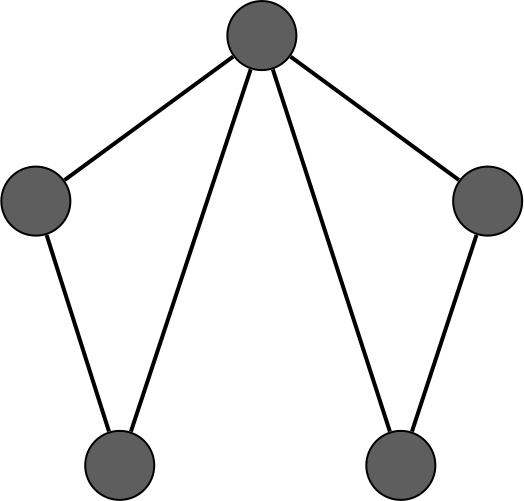}\vspace{-.3cm}
  \caption*{$H_{11}^{(5)}$}
\endminipage\hfill
\minipage{0.08\textwidth}
  \includegraphics[width=\linewidth]{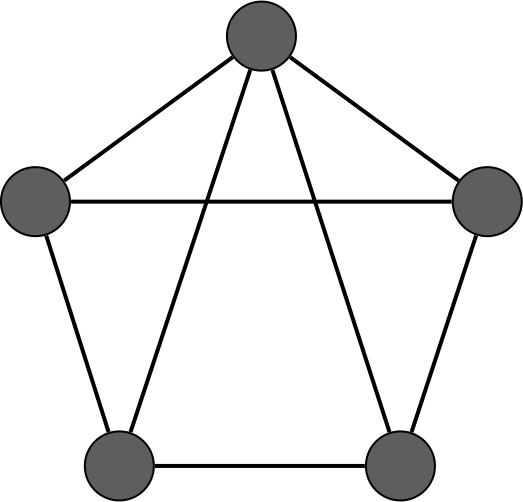}\vspace{-.3cm}
  \caption*{$H_{19}^{(5)}$}
\endminipage\hfill
\minipage{0.08\textwidth}%
  \includegraphics[width=\linewidth]{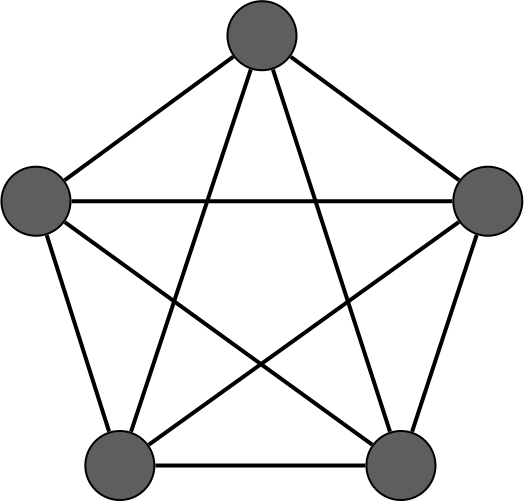}\vspace{-.3cm}
  \caption*{$H_{21}^{(5)}$}
\endminipage\hspace*{\fill}
\vspace{-.1cm}
\caption{Examples of graphlets}
\vspace{-.1cm}
\label{fig:graphlets}
\end{center}
\end{figure}

	A graphlet is a small connected graph topology, such as a triangle, wedge, or $k$-clique, which we will use to describe the local behavior of a larger network (example graphlets of size 3, 4, and 5, can be seen in Figure \ref{fig:graphlets}).
	Let the graph in question be $G = (V,E)$ where $V$ is a set of vertices and $E$ is a set of unordered pairs of vertices ($G$ is assumed to be connected, undirected, and unweighted).
	Imagine specifying a $k$-graphlet and testing for every induced subgraph of the graph (denoted $G|\{v_1,\ldots,v_k\}$ where $v_1,\ldots,v_k \in V$), if it is isomorphic to the subgraph (it has the same topology).
	We would like to compute the number of Connected Induced Subgraphs of size $k$ (denoted by {\em $k$-CIS} throughout) for which this match holds.
	We call this number the {\em graphlet counts} and the proportion of the number of such matches to the total number of $k$-CISs is called the {\em graphlet coefficient}.
	
    Graphlets are the graph analogue of wavelets (small oscillatory functions that are convolved with a signal to produce wavelet coefficients) because they are small topologies that are matched to induced subgraphs of the original graph to produce the graphlet coefficients.
	Graphlet coefficients, also referred to as graph moments, are used to fit certain graph models by the method of moments, \cite{bickel2011method}, and also are used to understand biological networks \cite{prvzulj2006efficient}.
	A naive graphlet counting method simply counts every induced subgraph which takes on the order of $n^k$ iterations.
	In a typical graph, the majority of induced subgraphs are disconnected, which would not count as a graphlet, so the majority of these iterates would not count toward the graphlet coefficient.
	We propose a class of Monte Carlo sampling methods called lifting that allow us to quickly estimate graphlet coefficients.
	The lifting step takes a CIS of size $k-1$ and produces a CIS of size $k$ by adding an adjacent vertex to it (according to a specific scheme), thereby forming graphlet samples in an inductive, bottom-up fashion (see~\ref{fig:lifting}).

	Monte Carlo sampling procedures perform random walks on graphlets of a certain size within a large network.
	These methods have the advantage of only requiring local graph information at every step, which makes them memory efficient in computation.
	The challenge in designing such an algorithm is showing that the sampling procedure is unbiased in its graphlet estimates, has low variance, and is sample efficient.
	Two such methods are GUISE algorithm of \cite{bhuiyan2012guise} and the pairwise subgraph random walk (PSRW) of \cite{Wang2014psrw}, which differ in the way they perform a random walk between CIS samples.
	Another option is to generate a sequence of vertices that induces a CIS sample, which has been done in \cite{Han2016waddling} using an algorithm called the Waddling random walk.
	Very efficient exact count methods exist \cite{Ahmed2017count, rahman2014graft, pinar2017escape, Bressan2017colourcoding},
	but they have not been extended to counting graphlets larger than $k=5$.

    We note that graphlet frequencies are one type of graph feature that relate to the proportion of motifs in a graph.
    However, they do not reflect more global properties of a graph, and are not comparable to graph embeddings such as GraphSAGE \cite{hamilton2017inductive} or node2vec \cite{grover2016node2vec}.
    Hence we do not offer such comparisons.

	\subsection{Our contributions}
    We provide two methods, the ordered lift estimator and the unordered lift estimator, which differ in the way that subgraphs are represented and counted.
    The ordered estimator allows for a modification, called {\em shotgun sampling} that samples multiple subgraphs in one shot, which effectively gives it more samples per iteration.
    For our theoretical component, we prove that the estimated graphlet coefficients are unbiased, and prove that the variance of the estimator scales like $\Delta^{k-2}$ where $\Delta$ is the maximum degree.
    We conclude with real-world network experiments that reinforce the contention that graphlet lifting is competitive with a specialized Waddling implementation and has better accuracy than subgraph random walks.
    We implement $6$-graphlet lifting on a 2.9M vertex Facebook graph, demonstrating that lifting is the first sampling scheme that can scale to 1M sized graphs and $k$-graphlets where $k > 5$, and do so without any specialized modifications.

	\section{Sampling graphlets}
	\subsection{Definitions and notation}
	Recall our definitions thus far: $G = (V,E)$ is a simple graph, $G|W$ is the induced subgraph for $W \subset V$.
	The set of all connected induced $k$-subgraphs (or $k$-CISs) of $G$ is denoted by $\cV_k(G)$ (or simply $\cV_k$).
	An unordered set of vertices is denoted $\{v_1,\ldots,v_k\}$ while an ordered list is denoted $[v_1,\ldots,v_k]$.
	Let $H_1, H_2, \ldots, H_l$ be all non-isomorphic
	motifs for which we would like the graphlet counts. 
	For $T\in \cV_k(G)$, we say that ``$T$ is subgraph of
	 type $m$'' if $T$ is isomorphic to $H_m$, and denote this with $T \sim H_m$.
	The number of $k$-subgraphs in $G$ of type $m$ is equal to 
	$N_m (G) = \sum_{T \in \cV_k(G)} \ind(T \sim H_m)$, where $\ind(A)$ 
	is the indicator function.
    For a subgraph $S \subseteq G$, denote $V_S$ to be the set of its vertices, $E_S$ to be the set of its edges.
	Denote $\cN_v(S)$ (vertex neighborhood of $S$) to be the set of all vertices adjacent to some vertex in $S$ not including $S$ itself. 
	Denote $\cN_e(S)$ (edge neighborhood of $S$) to be the set of all edges that connect a vertex from $S$ and a vertex outside of $S$.
	Also, denote $\deg(S)$ (degree of $S$) to be the number of edges in $\cN_e(S)$, and denote $\deg_S(u)$ ($S$-degree of $u$) to be the number of vertices from $S$ that are connected to $u$.
	Note that $\deg(S) + 2|E_S| = \sum_{v\in V_S} \deg(v)$.

    \subsection{Prior graphlet sampling methods}
	
The ideal Monte Carlo procedure would sequentially sample CISs uniformly at random from the set $\cV_k(G)$, classify their type, and update the corresponding counts.  
Unfortunately, uniformly sampling CISs is not a simple task because a random set of $k$ vertices is unlikely to be connected.
CIS sampling methods require Monte Carlo Markov Chains (MCMCs) for which one can calculate the stationary distribution, $\pi$, over the elements of $\cV_k$.
First, let us consider how we update the graphlet counts, $N_m(G)$, given a sample of CISs, $T_1, T_2, \ldots, T_n$.
Then we use Horvitz-Thompson inverse probability weighting to estimate the graphlet counts,
\begin{equation}
  \label{eq:moments}
  \hat N_m(G) := \frac{1}{n} \sum_{i=1}^n \frac{\ind(T_i\sim H_m)}{\pi(T_i)}.  
\end{equation}
It is simple to see that this is an unbiased estimate of the graphlet counts as long as $\pi$ is supported over all elements of $\cV_k$.

Let us describe the subgraph random walk in \cite{Wang2014psrw} called the pairwise subgraph random walk (PSRW).
In order to perform a random walk where the states are subgraphs $\cV_k$, we form the CIS-relationship graph.
Two $k$-CISs, $T,S \in \cV_k$ are connected with an edge if and only if vertex sets of $T$ and $S$ differ by one element, i.e. when $|V(T) \cap V(S)| = k-1$.
Given the graph structure, we sample $k$-CISs by a random walk on the set $\cV_k$, which is called Subgraph Random Walk (SRW).
Because the transition from state $S \in \cV_k$ is made uniformly at random to each adjacent CIS, we know that the stationary distribution will sample each edge in the CIS-relationship graph with equal probability.
This fact enables \cite{Wang2014psrw} to provide a local estimator of the stationary probability $\pi(S)$.
PSRW is a modification of the SRW algorithm, where each transition is performed from $S$ to $T$ in $\cV_{k-1}$ and then the $k$-CIS $S \cup T$ is returned.

Being a random walk-based procedure, insufficient mixing can cause PSRW to be biased if the burn-in period is not long enough.
It was pointed out in \cite{Bressan2017colourcoding} that the mixing time of the SRW can be of order $O(n^{k-2})$, even if the mixing time of the random walk on the original graph $G$ is of constant order $O(1)$.
PSRW also requires global constants based on the CIS-graph, which can be computationally intractable (super-linear time).
It should also be noted that a burn-in period is required for PSRW to converge to the stationary distribution, so any distributed sampling scheme will require all runs to perform this burn-in.

A naive method for sampling CIS's would be to perform a random walk on the graph, $G$, and then sample the $k$ vertices most recently visited.
This scheme is appealing because it has an easy to compute stationary distribution, and can `inherit' the mixing rate from the random walk on $G$ (which is relatively small).
Despite these advantages, certain graphlet topologies, such as stars, will never be sampled, and modifications are needed to remedy this defect.
\cite{chen2016general} combined this basic idea with the SRW by maintaining a $l$ length history of the SRW on CISs of size $k-l+1$, and unioning the history, but this suffers from the same issues as SRW, such as slow mixing and the need to calculate global constants based on the CIS-graph.

\cite{Han2016waddling} introduced a {\em Waddling} protocol which retains a memory of the last $s$ vertices in the random walk on $G$ and then extends this subgraph by $k-s$ vertices from either the first or last vertex visited in the $s$-subgraph (this extension is known as the `waddle').
Waddling requires that one samples from the stationary distribution over $G$, but this can be achieved by selecting an edge uniformly at random from the graph, thus avoiding the burn-in.
The authors provide recommendations for calculating the stationary distribution for this MCMC, and prove a bound on the error for the graphlet coefficients.
The upside to this method is that the precise Waddling protocol used should depend on the desired graphlet, and the algorithm involves a rejection step which may lead to a loss of efficiency.
This is simultaneously a downside of the method: the general specification of the method makes the algorithm implementation difficult.
In contrast, lifting requires little tuning,  perhaps at the expense of customizability.
Finally, lifting has the advantage of never rejecting graphlets, has similar theoretical guarantees, and has simple parallel extensions.

\section{Subgraph lifting}
	
	The lifting algorithm is based on a randomized protocol of attaching a vertex to a given CIS.
	For any \smash{$(k-1)$}-CIS, $S$, we lift it to a $k$-subgraph by adding a vertex from its neighborhood, $\cN_v(S)$ at random according to some probability distribution.
	Note that this basic lifting operation can explore any possible subgraph in $\cV_k$. 

\begin{figure}[th]
\begin{center}
\hspace*{\fill}
\minipage{0.10\textwidth}
  \includegraphics[width=\linewidth]{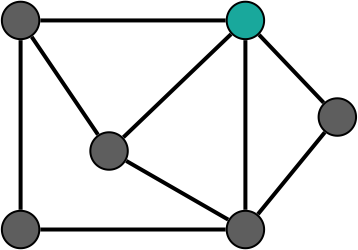}
  \caption*{(a)}
\endminipage\hfill
\minipage{0.10\textwidth}
  \includegraphics[width=\linewidth]{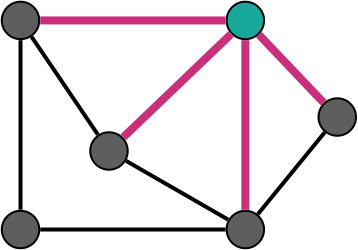}
  \caption*{(b)}
\endminipage\hfill
\minipage{0.10\textwidth}
  \includegraphics[width=\linewidth]{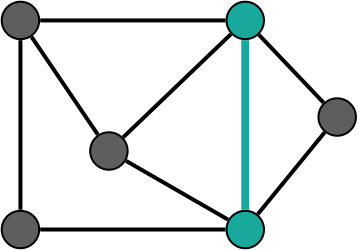}
  \caption*{(c)}
\endminipage\hfill
\minipage{0.10\textwidth}%
  \includegraphics[width=\linewidth]{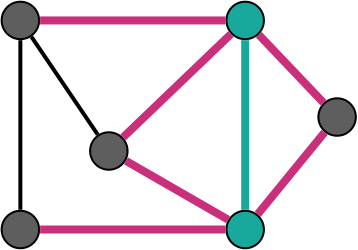}
  \caption*{(d)}
\endminipage
\hspace*{\fill}
\vskip 5pt
\hspace*{\fill}
\minipage{0.10\textwidth}
  \includegraphics[width=\linewidth]{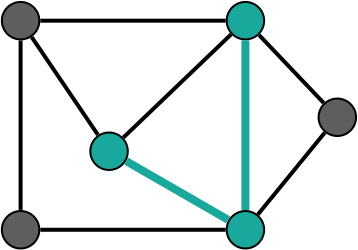}
  \caption*{(e)}
\endminipage\hfill
\minipage{0.10\textwidth}
  \includegraphics[width=\linewidth]{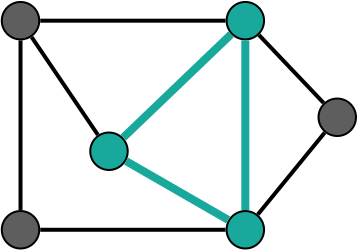}
  \caption*{(f)}
\endminipage\hfill
\minipage{0.10\textwidth}
  \includegraphics[width=\linewidth]{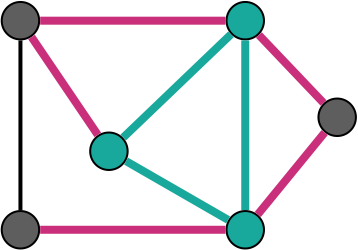}
  \caption*{(g)}
\endminipage\hfill
\minipage{0.10\textwidth}%
  \includegraphics[width=\linewidth]{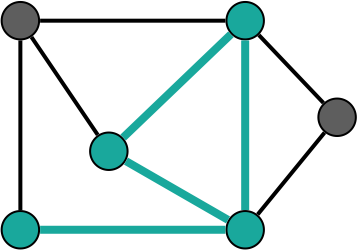}
  \caption*{(h)}
\endminipage\hspace*{\fill}
\caption{Lifting procedure}
\label{fig:lifting}
\end{center}
\end{figure}

    You can see an example of the lifting sampling scheme in Figure \ref{fig:lifting}, where the algorithm iteratively builds a $4$-CIS from a chosen node.
	First assume we have a node $v_1$ sampled from the distribution $\pi_1$, a base distribution that can be computed from local information (step (a)). 
	We assume that $\pi_1(v) = \frac{f(\deg(v))}{K}$, where $f(x)$ is some function (usually a polynomial) and $K$ is some global normalizing constant which is assumed to be precomputed.
	Denote $S_1 = \{v_1\}$.
    To start our procedure, sample an edge $(v_1,v_2)$ uniformly from $\cN_e(S_1)$ (step (b)). 
	The vertex $v_2$ is then attached to $S_1$, forming a subgraph $S_2 = G|(V_{S_1} + {v_2})$ (step (c)).
	After that, we sample another edge $(v_i, v_3)$ (with $1\leq i\leq 2$) uniformly from $\cN_e(S_2)$, and the vertex $v_3$ is then attached to $S_2$ (steps (d-f)).
	At each step we sample an edge $(v_i, v_{r+1})$ (with $1\leq i \leq r$) from $\cN_e(S_r)$ uniformly at random, and attach the vertex $v_{r+1}$ to the subgraph $S_r$ (steps (g-h)).
	After $k-1$ operations, we obtain a $k$-CIS, $T = S_k$.
	We'll refer to the procedure above as the {\em lifting procedure starting at vertex $v_1$}.

    Once a $k$-CIS, $T$, has been sampled we need to classify its graphlet topology, $H_m \sim T$.
    Because lifting does not target specific graphlet topologies, we need to be prepared to modify the coefficient for any graphlet (the coefficients are elaborated on in the next section).

    By induction, we can see that every $k$-CIS has a non-zero probability of being visited, assuming that $\pi_1$ is supported on every vertex.
    We consider two options for the starting vertex, $\pi_1$: uniform distribution over vertices, and the stationary distribution for a simple random walk on $G$. 
	Lifting and waddling both can `inherit' the mixing time of a simple random walk by initializing with the stationary distribution.
	In addition, lifting can be parallelized by having each thread start at a random vertex uniformly, while waddling requires us to start from the stationary distribution.
	In the next section, we show how to calculate the probability of sampling the $k$-CIS, $\pi(S)$, using only its local information.

	\subsection{Unordered lift estimator}
	We can recursively compute the marginal probability of sampling the graphlet, $\pi_U(T)$, for the lifted CIS $T \in \cV_k(G)$.
	We say that this method is unordered because we ignore the order in which we visit the vertices in the graphlet.
    One advantage of this approach is that this probability is a function of only the degrees of vertices $V_T$.
	This can be done either recursively or directly. 
	Throughout, let the set of vertices of $T$ be $v_1,\ldots,v_k$.
	
	We begin the algorithm by querying the probability of obtaining any vertex in $T$, $\pi_1(v_i), i=1,\ldots,k$.
	We will build the probability of obtaining any connected subgraph of $T$ inductively.
	This is possible because the probability of getting $T$ via lifting is given by the sum $\pi_U(T) = \sum_S \mathbb{P}(T|S) \pi_{U}(S)$, where the sum is taken over all connected \smash{$(k-1)$}-subgraphs $S\subset T$, and $\mathbb{P}(T|S)$ denotes the probability of getting from $S$ to $T$ in the lifting procedure.
	Then
	\begin{align}
	\label{eq:recursive}
		\pi_{U}(T) & = \sum_{S\subset T} \pi_{U}(S) \frac{\deg_S (V_T\setminus V_S)}{|\cN_e(S)|} \\
		& = \sum_{S\subset T} \pi_{U}(S) \frac{|E_T| - |E_S|}
		{\sum_{u\in S} \deg(u) - 2|E_S|}, \nonumber
	\end{align}
	where the sum is taken over all connected \smash{$(k-1)$}-subgraphs $S\subset T$.
	
	Consider the sampled $k$-CIS $T := S_k$.
	Denote the set of possible sequences $A = [v_1,\ldots, v_k]$ that would form $T$ in the lifting process as $\co(T)$.
	Notice that $S_r = G|\{v_1,\ldots,v_r\}$ must be a connected subgraph for all $r$.
	Thus,
	\begin{multline}
	\label{eq:co}
	    \co(T) = \big\{\,[v_1,\ldots,v_k]\in V^k_G \mid \{v_1,\ldots,v_k\} = V_T,\\  T|\{v_1,\ldots,v_r\} \text{ is connected }\big\}.
	\end{multline}
	Since the elements of $\co(T)$ are just certain orderings of vertices in $T$, we call an element from $\co(T)$ a \textit{compatible ordering} of $T$.
	Note that $|\co(T)|$ only depends on the type of the graphlet isomorphic to $T$, and it can be precomputed using dynamic programming.
	Thus, when $T\sim H_m$, the number of compatible orderings are equal: $|\co(H_m)| = |\co(T)|$. Note that $|\co(H_m)|$ can vary from $2^{k-1}$ (for $k$-path) to $k!$ (for $k$-clique).
	For a direct formula, we notice that $\pi_U(T) = \sum_{A\in \co(T)} \tilde\pi(A)$, and $\tilde\pi(A)$ is the probability of getting sequence $A\in \co(T)$ in the lifting process (see \eqref{eq:co},\eqref{eq:tildepi}).
	Then 
	\begin{equation}
	\label{eq:direct}
		\pi_U(T) = \sum_{A\in\co(T)} \frac{f(\deg(A[1]))}{K} \prod_{r=1}^{k-1} \frac{|E_{S_{r+1}(A)}| - |E_{S_r(A)}|}{\sum_{i=1}^r \deg(A[i]) - 2|E_{S_r(A)}|},
	\end{equation}
	where, given $A=[v_1,\ldots,v_k]$, $A[i]$ is the $i$th vertex in $A$ and $S_r(A) = G|\{v_1,\ldots,v_r\}$.
	
	Although calculation of this probability on-the-fly is cost-prohibitive, we can greatly reduce the number of operations by noticing that the probability $\pi_k(T)$ is a 
	function of degrees of the vertices: for a CIS $T$ of type $m$, let $[v_1, \ldots, v_k]$ be an arbitrary labelling of the vertices of $T$ with $d_i = \deg(v_i)$, then the probability of $T$ is
	\begin{equation*}
	    \pi_U(T) =\frac{1}{K} F_m(d_1, \ldots, d_k)
	\end{equation*}
    for a cached function $F_m$ given by \eqref{eq:direct}.
	
	\textbf{Example.} Consider a triangle, which is a 3-graphlet with edges $(v_1,v_2)$, 
	$(v_2, v_3)$ and $(v_1,v_3)$. Given the degrees $d_1, d_2, d_3$ of the corresponding 
	vertices, the probability function is
	\begin{align}
	\label{prob:triangle}
		\pi_U(\mathrm{triangle}) & = \left( \frac{\pi_1(d_1)}{d_1} + 
		\frac{\pi_1(d_2)}{d_2}\right) \frac{2}{d_1+d_2-2} \nonumber \\ 
		& + \left( \frac{\pi_1(d_2)}{d_2} + \frac{\pi_1(d_3)}{d_3}\right) 
		\frac{2}{d_2+d_3-2} \nonumber \\
		& + \left( \frac{\pi_1(d_3)}{d_3} + \frac{\pi_1(d_1)}{d_1}\right) 
		\frac{2}{d_3+d_1-2}.
	\end{align}
	
	\textbf{Example.} Consider a wedge, which is a 3-graphlet with edges $(v_1,v_2)$ and 
	$(v_1,v_3)$. Given the degrees $d_1, d_2, d_3$ of the corresponding 
	vertices, the probability function is
	\begin{align}
	\label{prob:wedge}
		\pi_U(\mathrm{wedge}) & = \left( \frac{\pi_1(d_1)}{d_1} + \frac{\pi_1(d_2)}{d_2}\right) \frac{1}{d_1+d_2-2} \nonumber \\
		& + \left( \frac{\pi_1(d_1)}{d_1} + \frac{\pi_1(d_3)}{d_3}\right) \frac{1}{d_1+d_3-2}.
	\end{align}

	We need to only compute functions $F_m$ once before starting the algorithm.
	When a $k$-CIS $T$ is sampled via lifting procedure, we find the natural labelling of vertices in $T$ via the isomorphism $H_m \rightarrow T$, and use the function $F_m$ together with the degrees $d_1,\ldots,d_k$ of vertices of $T$ to compute the value of 
	$\pi_U(T) = \frac{1}{K} F_{m}(d_1,\ldots,d_k)$.
	
	
\begin{algorithm}[h]
\label{alg:ULE}
\caption{Unordered Lift Estimator}
\begin{algorithmic}
    \INPUT Graph $G$, graphlet size $k$
    \OUTPUT $\hat N_m(G)$
    \STATE For each $k$-graphlet in canonical form, $H_m$, precompute the function $F_m(d_1,\ldots, d_k)$ and the global constant $K$
    \STATE Initialize $v$ at an arbitrary node, $n \gets 0$, $\hat N_m(G) \gets 0$
    \WHILE{stopping criteria is not met}
        \STATE Sample initial vertex $v$ from $\pi_1(v)$
        \STATE Initialize $V_T \gets \{v\}$ and $E_T \gets \{\}$
        \STATE Initialize $\cN_e(T)\gets \cN_e(v)$
        \WHILE{$|V_T| < k$}
            \STATE Sample an edge $e=(v,u)$ uniformly from $\cN_e(T)$, with $v\in V_T$ and $u\notin V_T$
            \STATE Set $E_T(u) \gets \{(v,u)\in \cN_e(T)\}$
            \STATE Update $V_T\gets V_T\cup\{u\}$ and $E_T \gets E_T \cup E_T(u)$
            \STATE Query $\cN_e(u)$
            \STATE Update $\cN_e(T) \gets [\cN_e(T)\cup\cN_e(u)]\setminus E_T(u)$
        \ENDWHILE
        \STATE Set $H_m = {\rm hash}(T)$
        \STATE Determine the ordering $[v_1,\ldots,v_k]$ of vertices in $V_T$ induced by the isomorphism $(V_T, E_T)\sim H_m$
        \STATE Set $d_i = |\cN_e(v_i)|$ for all $i=1,\ldots, k$
        \STATE Set $\pi(T) = \frac{1}{K}F_m(d_1,\ldots,d_k)$
        \STATE Update $\hat N_m(G) \gets \hat N_m(G) + \pi^{-1}(T)$
       \STATE Update $n \gets n + 1$
    \ENDWHILE
    \STATE Normalize $\hat N_m(G) \gets \frac{1}{n}\hat N_m(G)$
\end{algorithmic}
\end{algorithm}

	\subsection{Ordered lift estimator}
	
    
    The sample estimator, \eqref{eq:moments}, does not track the order of the vertices as they are sampled to form a graphlet.
    We can, however, track the vertex information and thus define an estimator on ordered sequences of vertices $[v_1, \ldots, v_k]$, denoted by $A$.
    Given a sampling scheme of such sequences with probability $\tilde\pi(A)$, the estimator for graphlet counts is given by
    \begin{equation}
      \label{eq:sequences}
      \hat N_m(G) := \frac{\omega_m}{n} \sum_{i=1}^n \frac{\ind(G|A_i\sim H_m)}{\tilde\pi(A_i)}
    \end{equation}
    for some fixed weights $\omega_m$.
    The main difference between these types of sampling is that we maintain the ordering of the vertices, while a CIS is an unordered set of vertices.
    
    We can think of a lifting procedure as a way of sampling a sequence $A = [v_1, \ldots, v_k]$, ordered from the first vertex sampled to the last, that is then used to generate a CIS.
	Denote the set of such sequences as $V^k_G$.
	Let $S_r = G|\{v_1,\ldots, v_r\}$ be the $r$-CIS obtained by the lifting procedure on step $r$.
	The probability of sampling vertex $v_{r+1}$ on the step $r+1$ is equal to
	\begin{equation*}
	    \mathbb{P}(v_{r+1}|S_{r}) := \frac{\deg_{S_r}(v_{r+1})}{|\cN_e(S_r)|} =
	    \frac{|E_{S_{r+1}}| - |E_{S_r}|}{\sum_{i=1}^r \deg(v_i) - 2|E_{S_r}|}.
	\end{equation*}
	Thus, the probability of sampling a sequence $A \in V^k_G$ is equal to
	\begin{align}
	\label{eq:tildepi}
	    \Tilde\pi(A) & \coloneqq \pi_1(v_1)\prod_{r=1}^{k-1} \P\left(v_{r+1}|S_r\right) \nonumber \\
	    & = \frac{f(\deg(v_1))}{K} \prod_{r=1}^{k-1} \frac{|E_{S_{r+1}}| - |E_{S_r}|}{\sum_{i=1}^r \deg(v_i) - 2|E_{S_r}|}.
	\end{align}
    Critically, this equation can be computed with only neighborhood information about the involved vertices, so it takes $O(k)$ neighborhood queries.
    Because there are many orderings that could have led to the same CIS $T$, then we need to apply proper weights in the graphlet count estimate \eqref{eq:sequences} by enumerating the number of possible orderings.
    
	We set up the estimator from \eqref{eq:sequences} as
	\begin{equation}
	\label{eq:ord_est}
	    \hat N_{O,m} := \frac{1}{n}\frac{1}{|\co(H_m)|} \sum_{i=1}^n \frac{\ind(G|A_i\sim H_m)}{\tilde\pi(A_i)}.
	\end{equation}
	We call it the {\em ordered lift estimator} for the graphlet count.

\begin{algorithm}[h]
\label{alg:OLE}
\caption{Ordered Lift Estimator (with optional shotgun sampling)}
\begin{algorithmic}
    \INPUT Graph $G$, graphlet size $k$
    \OUTPUT $\hat N_m(G)$
    \STATE Count $|\co(H_m)|$- the number of compatible orderings in $H_m$.
    \STATE Initialize $v$ at an arbitrary node, $n \gets 0$, $\hat N_m(G) \gets 0$
    \WHILE{stopping criteria is not met}
        \STATE Sample $v_1$ from $\pi_1(v)$
        \STATE Initialize $V_S \gets \{v_1\}$ and $E_S \gets \{\}$
        \STATE Initialize $\cN_e(S)\gets \cN_e(v_1)$
        \STATE Initialize $\pi(S) \gets \pi_1(v_1)$
        \WHILE{$|V_S| < k-1$}
            \STATE Sample an edge $e=(v,u)$ uniformly from $\cN_e(S)$, with $v\in V_S$ and $u\notin V_S$
            \STATE Set $E_S(u) \gets \{(v,u)\in \cN_e(S)\}$
            \STATE Update $\pi(S)\gets \pi(S)\frac{|E_S(u)|}{|\cN_e(S)|}$
            \STATE Update $V_S\gets V_S\cup\{u\}$ and $E_S \gets E_S \cup E_S(u)$
            \STATE Query $\cN_e(u)$
            \STATE Update $\cN_e(S) \gets [\cN_e(S)\cup\cN_e(u)]\setminus E_S(u)$
        \ENDWHILE
        \IF {not shotgun sampling}
            \STATE Sample an edge $e=(v,u)$ uniformly from $\cN_e(S)$, with $v\in V_S$ and $u\notin V_S$
            \STATE Set $E_S(u) \gets \{(v,u)\in \cN_e(S)\}$
            \STATE Set $\pi(T)\gets \pi(S)\frac{|E_S(u)|}{|\cN_e(S)|}$
            \STATE Set $V_T\gets V_S\cup\{u\}$ and $E_T \gets E_S\cup E_S(u)$
            \STATE Set $H_m = {\rm hash}(T)$
            \STATE Update $\hat N_m(G) \gets \hat N_m(G) + \pi^{-1}(T)$ 
        \ENDIF
        \IF {shotgun sampling}
            \FORALL{$u\in \cN_v(S)$}
                \STATE Set $E_S(u) \gets \{(v,u)\in \cN_e(S)\}$
                \STATE Set $V_T\gets V_S\cup\{u\}$ and $E_T \gets E_S \cup E_S(u)$
                \STATE Set $H_m = {\rm hash}(T)$
                \STATE Update $\hat N_m(G) \gets \hat N_m(G) + \pi^{-1}(S)$
            \ENDFOR
        \ENDIF
        \STATE Update $n \gets n + 1$
    \ENDWHILE
    \STATE Normalize $\hat N_m(G) \gets \frac{1}{n}\frac{1}{|\co(H_m)|}\hat N_m(G)$
\end{algorithmic}
\end{algorithm}
	
	A drawback of the algorithm is that it takes $k-1$ queries to lift the CIS plus the number of steps required to sample the first vertex (when sampled from Markov chain). 
	To increase the number of samples per query, notice that if we sample $B = [v_1, \ldots, v_{k-1}]$ via lifting, we can get subgraphs induced by $A = [v_1,\ldots, v_{k-1},u]$ for all $u \in \cN_v(B)$ without any additional queries.
	
	Thus, for each sampled sequence $B_i \in V_G^{k-1}$, we can compute the sum $\sum_{u\in \cN_v(B_i)} \ind(G|B_i \cup \{u\} \sim H_m)$ to incorporate the information about all $k$-CISs in the neighborhood of $B_i$.
	We call this procedure {\em shotgun sampling}.
	The corresponding estimator based on \eqref{eq:sequences} is
	\begin{equation}
	\label{eq:shot_est}
	    \hat N_{S,m} = \frac{1}{n}\frac{1}{|\co(H_m)|} \sum_{i=1}^n \frac{\sum_{u \in \cN_v(B_i)} \ind(G|B_i \cup \{u\} \sim H_m)}{\tilde \pi(B_i)}.
	\end{equation}
	Shotgun sampling produces more CIS samples with no additional query cost, but the CIS samples generated in a single iteration will be highly dependent.
	The following proposition states that the resulting estimators are unbiased (see Appendix for the proof).
	
	\begin{proposition}
	\label{prop:unbiased}
    The ordered lifted estimator, $\hat N_{O,m}$, and the shotgun estimator, $\hat N_{S,m}$, are unbiased for the graphlet counts $N_m$.
    \end{proposition}
	

	\section{Lifting Variance}
	One advantage of the lifting protocol is that it can be decoupled from the selection of a starting vertex, and our calculations remained agnostic to the distribution $\pi_1$ (although, we did require that it was a function of the degrees).
    There are two methods that we would like to consider: one is the uniform selection over the set of vertices and the other is from a random walk on the vertices, that presumably has reached its stationary distribution. 
	
	Consider sampling the starting vertex $v$ independently and from an arbitrary distribution $\pi_1$ when we have access to all the vertices.
    The advantage of sampling vertices independently, is that the lifting process will result in independent CIS samples.
    A byproduct of this is that the variance of the graphlet count estimator \eqref{eq:moments} can be decomposed into the variance of the individual CIS samples.
    Given iid draws, the variance of the estimator $\hat N_m(G)$ is then  
	\begin{align}
	\label{eq:variation.ind}
	    V_{m}^{\independent}(\hat N_{U,m}) & \coloneqq \frac 1n \Var\left( \frac{\ind(T_n \sim H_m)}{\pi_U(T_n)} \right) \nonumber \\
	    & = \frac{1}{n}\left(\sum_{T \in \cV_k} \frac{\ind(T\sim H_m)}{\pi_U(T)} - N_m(G)^2\right),
	\end{align}
	which is small when the distribution of  $\pi_U(T)$ is close to uniform distribution on $\cV_m(G)$.
	Equation \eqref{eq:variation.ind} demonstrates fundamental property that when $\pi_U(T)$ is small then it contributes more to the variance of the estimator.
	The variation in \eqref{eq:variation.ind} can be reduced by an appropriate choice of $\pi_1$, i.e.~the starting distribution.
	
	For example, if $k=3$, let $\pi_1(v) = \frac{1}{K}\deg(v)(\deg(v)-1)$, where 
	$K=\sum_{u\in V_G} \deg(u)(\deg(u)-1)$. 
	Then by \eqref{prob:triangle} and \eqref{prob:wedge}
	\begin{equation*}
		\pi_U(\mathrm{triangle}) = \frac{6}{K}, \quad \pi_U(\mathrm{wedge}) = \frac{2}{K}.
	\end{equation*}
	Calculating $K$ takes $O(|V_G|)$ operations (preparation), sampling starting vertex $v$ takes $O(\log(|V_G|))$ operations, and lifting takes $O(\Delta)$, where $\Delta$ is the maximum vertex degree in $G$.

	When we don't have access to the whole graph structure, a natural choice is to run a simple random walk (with transitional probabilities \smash{$p(i \small \to j) = \frac{1}{\deg(i)}$} whenever $j$ in connected to $i$ with an edge).
	Then the stationary distribution is $\pi_1(v) = \deg(v) / (2|E_G|),$ and we can calculate all probabilities $\pi_k$ accordingly.
	One feature of the simple random walk is that the resulting edge distribution is uniform: $\pi_U(e) = \frac{1}{|E_G|}$ for all $e\in E_G$ (edges are $2$-graphlets). 
	Therefore, the probabilities $\pi_U$ are the same as if sampling an edge uniformly at random and start Lifting procedure from that edge.

\subsection{Theoretical variance bound}

As long as the base vertex distribution, $\pi_1$, is accurate then we have that the graphlet counts are unbiased for each of the aforementioned methods.
The variance of the graphlet counts will differ between these methods and other competing algorithms such as Waddling and PSRW.
The variance of sampling algorithms can be decomposed into two parts, an independent sample variance component and a between sample covariance component.
As we have seen the independent variance component is based on the properties of $\pi$ resulting from the procedure (see \eqref{eq:variation.ind}).
We have three different estimators: Ordered Lift estimator $\hat N_{O,m}$, Shotgun Lift estimator $\hat N_{S,m}$ and Unordered Lift estimator $\hat N_{U,m}$.
For each estimator, we sample different objects: sequences $A_i \in V_G^k$ for Ordered, sequences $B_i \in V_G^{k-1}$ for Shotgun, and CISs $T_i\in \cV_k(G)$ for Unordered estimator.
Throughout this section, we will denote 
\begin{enumerate}
    \item for the Ordered Lift estimator,
    \begin{equation}
    \label{phi:ord}
        \phi_{O,i} = \frac{\ind(G|A_i\sim H_m)}{|\co(H_m)|\tilde\pi(A_i)},
    \end{equation} 
    
    \item for the Shotgun Lift estimator,
    \begin{equation}
    \label{phi:shot}
        \phi_{S,i} =  \frac{\sum_{u \in \cN_v(B_i)} \ind(G|B_i \cup \{u\} \sim H_m)}{|\co(H_m)| \tilde \pi(B_i)},
    \end{equation}
    
    \item for the Unordered Lift estimator,
    \begin{equation}
    \label{phi:unord}
        \phi_{U,i} = \frac{\ind(T_i\sim H_m)}{\pi_U(T_i)}.
    \end{equation}
    
\end{enumerate}
Let $\phi_1$ be shorthand for $\phi_{X,1}$, where $X \in \{O,S,U\}$, and note that $N_m(G) = \E \phi_1$, and $\hat N_m(G) = \frac{1}{n}\sum_i \phi_i$ for the corresponding estimators.


The variance can be decomposed into the independent sample variance and a covariance term,
\begin{equation}
\label{eq:variance}
\Var(\hat N_m(G)) = \frac{1}{n}V_m^{\independent}(\phi_1) + \frac{2}{n^2} \sum_{i < j} \mathbb \Cov \left( \phi_i, \phi_j \right).
\end{equation}
For Markov chains, the summand in the second term will typically decrease exponentially as the lag $j-i$ increases, due to mixing.
If we start from a random vertex then the samples are uncorrelated and the covariance term disappears.
For an analysis of the mixing time for random walk-based graphlet Lifting, see the Appendix.

Let us focus on the first term, with the goal of controlling this for either choice of base vertex distribution, $\pi_1$, and the lifting scheme.

\begin{theorem}
  \label{thm:var_bd}
  Let $\phi_1$ be as defined in \eqref{phi:ord}, \eqref{phi:shot} or \eqref{phi:unord}.
  Denote the first $k$ highest degrees of vertices in $G$ as $\Delta_1,\ldots, \Delta_k$ and denote $D = \prod_{r=2}^{k-1} (\Delta_1 +\ldots + \Delta_r)$.
  
  (1) If $\pi_1$ is the stationary distribution of the vertex random walk then
    \begin{equation}
    V_m^{\independent}(\phi_1) \leq N_m(G) \frac{2|E_G|}{|\co(H_m)|} D.
\end{equation}
  (2) If $\pi_1$ is the uniform distribution over the vertices then 
  \begin{equation}
      V_m^{\independent}(\phi_1) \leq N_m(G) \frac{2 \Delta_1 |E_G|}{|\co(H_m)|} D.
  \end{equation}
\end{theorem}
This result is comparable to analogous theorems for Waddling, \cite{Han2016waddling}, and PSRW, \cite{Wang2014psrw}. 
Critically, Lifting works without modification for all graphlets up to a certain size.
It should be noted that the variance of each lift method has the same bound in Theorem \ref{thm:var_bd}. 
We do not observe significant differences between the empirical variances of the unordered and ordered lifts.
The shotgun method does significantly reduce the observed variance, because it samples more graphlets per iteration, but due to the dependence between samples within a single lift, this is not reflected in the theory.

\section{Experiments}
\subsection{Description of experiments}
All experiments were implemented on Amazon Web Services `t2.xlarge' instances running Ubuntu 16.04 (January 2019).
All algorithms were implemented in Python, the code for which is available on GitHub\footnote{github.com/dshemetov/GraphletLift}.
Throughout our experiments we only compare against graphlet Monte Carlo sampling algorithms and do not compare against exact graphlet counting methods (except in computing a ground truth).
This is consistent with our thesis, that Lifting can accurately compute graphlet coefficients with a moderate number of samples that only require neighborhood look-ups (as opposed to processing the whole graph and counting all graphlets).

We implemented our own Waddle and PSRW protocols, for clean comparisons.
To get true count values, we used ESCAPE \cite{pinar2017escape} for $k=5$ and PGD \cite{Ahmed2017count} for $k=3,4$.
All the methods were studied under the same number of iterations where they had comparable run times.
The ground truth algorithms, ESCAPE and PGD, were faster than our estimation method, but these methods are limited to $k < 6$; we are aware of no exact counting method that does not hit the hard complexity barrier for large graphlet counts.

The Lifting method for $k$ graphlets was implemented as follows.
The initialization proceeds by pre-computing the probability functions $F_m$ for every graphlet in the atlas of graphlets of size $k$ and caching them symbolically through SymPy.
The probability functions, $\pi$, are stored in a dictionary keyed by a canonical graph labeling string certificate generated by \textit{nauty}~\cite{mckay2014practical} to reduce the cost of graph isomorphism checks.
In every iteration of Lifting we: sample a random node, lift up to a $k$-node graphlet, get the cached probability function $F_m$ by graph hashing, and, finally, find an isomorphism between the sampled graph and the canonical graph to obtain the probability of sampling the graphlet. 
Summing the inverses of these probabilities gives the estimate.

For our experiments, we picked five networks of different size, density, and domain ~\cite{nr-aaai15}\footnote{Network names correspond online datasets at networkrepository.com}.
The size of the graphs is listed in Table~\ref{tab:data_sets_info}.
\begin{itemize}
    \item The CELE network is a list of edges of the metabolic network of \textit{C. elegans}. 
    \item The EMAIL network is a university email exchange network.
    \item The CAIDA network is a network of packet routing relationships between AS's (e.g. Internet Service Providers).
    \item The FULLB network corresponds to a large positive definite matrix arising from a finite-element method.
    \item The SOCFB network is a network of user friendships on Facebook circa September 2005.
\end{itemize}

\begin{figure}
\begin{center}
    \begin{tabular}{ | l || r | r | r |}
        \hline
        Network name & $|V_G|$ & $|E_G|$ & Avg. Deg. \\ \hline
        \textbf{bio-celegansneural} (CELE) & 297 & 2,148 & 15 \\
        \textbf{ia-email-univ} (EMAIL) & 1,133 & 5,451 & 9 \\
        \textbf{misc-as-caida} (CAIDA) & 26,475 & 52,281 & 1.97 \\
        \textbf{misc-fullb} (FULLB) & 199,187 & 5.7M & 28.9 \\
        \textbf{socfb-B-anon} (SOCFB) & 2.9M & 20.9M & 14 \\
        \hline
    \end{tabular}
\end{center}
\caption{Networks used in experiments (M = millions).} 
\label{tab:data_sets_info}
\end{figure}

\subsection{Comparisons on $4$-graphlets}
We performed a full comparison over all $4$-graphlets (6 topologies), all networks (5 datasets), and three methods (unordered lift, PSRW, Waddle).
Using the relative error between the estimate $\hat N_m$ and the ground truth $N_m$ defined by
\begin{equation*}
    \text{Relative Error} = \frac{|\hat N_m(G) - N_m(G)|}{N_m(G)},
\end{equation*}
we can compare the performance of the algorithms on estimating each graphlet.
Fixing iterations to 40K, we produced the relative errors for the algorithms across all graphs and all $4$-graphlets in Figure \ref{tab:relative_errors}.
On the CELE graph, lifting outperforms on all graphlets.
On the EMAIL graph, PSRW rivals lifting on some of the graphlets.
Lifting has its worst performance on the CAIDA dataset, which the authors suspect is because the graph is extremely sparse and is mostly stars; rare graphlets, such as $H_6^{(4)}$, are difficult to detect for all methods.
However, lifting is only the worst of the three methods on the 3-star graph for CAIDA.
On the plus-side, lifting demonstrates the ability to find rare graphlets in large graphs, such as $H_4^{(4)}, H_5^{(4)}, H_6^{(4)}$ in SOCFB.

To get a sense for the convergence rates, we can plot the convergence to the true count as a function of iterations. 
We show this in 
Figure \ref{fig:converge_to_error} for the $4$-graphlets on the FULLB graph.
Overall, we find comparable performance among the three algorithms on the 3-star, 4-tailed triangle, and the 4-clique.
In some cases, such as $H_5^{(4)}$ and $H_4^{(4)}$, PSRW does not converge to the truth in the allotted number samples.
This may be due to the mixing rate of PSRW, which was not fast enough, leading to bias in the estimated sampling probability.
Waddle and lifting do approximately equally well on all the graphlets.

\begin{figure}
\begin{center}
        \begin{tabular}{ | l | l | r || r | r | r | r | r |}
        \hline
        \multicolumn{3}{|c||}{Network/Graphlet} & \multicolumn{3}{|c|}{Relative Error}\\ \hline
        Network & Graphlet & Freq & Lift & PSRW & Waddle \\ \hline
        \textbf{CELE} & $H_1^{(4)}$ & 0.4668 & \textbf{0.0075} & 0.0180 & 0.2153 \\
        & $H_2^{(4)}$ & 0.3703 & \textbf{0.0024} & 0.0301 & 0.1938 \\
        & $H_3^{(4)}$ & 0.1336 & \textbf{0.0118} & 0.0225 & 0.2055 \\
        & $H_4^{(4)}$ & 0.0113 & \textbf{0.0063} & 0.3241 & 0.1802 \\
        & $H_5^{(4)}$ & 0.0163 & \textbf{0.0079} & 0.1184 & 0.1978 \\
        & $H_6^{(4)}$ & 0.0014 & \textbf{0.0077} & 0.0865 & 0.1831 \\
        \hline
        \textbf{EMAIL} & $H_1^{(4)}$ & 0.2865 & \textbf{0.0009} & 0.0083 & 0.1934 \\
        & $H_2^{(4)}$ & 0.5803 & 0.0062 & \textbf{0.0014} & 0.1587 \\
        & $H_3^{(4)}$ & 0.1137 & \textbf{0.0058} & \textbf{0.0058} & 0.2049 \\
        & $H_4^{(4)}$ & 0.0066 & \textbf{0.0462} & 0.3213 & 0.1585 \\
        & $H_5^{(4)}$ & 0.0108  & \textbf{0.0239} & 0.1656 & 0.2134 \\
        & $H_6^{(4)}$ & 0.0017 & 0.0498 & \textbf{0.0369} & 0.1113 \\
        \hline
        \textbf{CAIDA} & $H_1^{(4)}$ & 0.9588 & 0.0313 & \textbf{0.0038} & 0.0132 \\
        & $H_2^{(4)}$ & 0.03505 & 0.0525 & 0.0891 & \textbf{0.0126} \\
        & $H_3^{(4)}$ & 0.0058 & 0.0774 & \textbf{0.0740} & 0.0883 \\
        & $H_4^{(4)}$ & 5e-05 & 0.0355 & 0.2219 & \textbf{0.0134} \\
        & $H_5^{(4)}$ & 0.0002 & \textbf{0.0039} & 0.6531 & 0.1996 \\
        & $H_6^{(4)}$ & 6.6e-06  & 0.6534 & 1.0000 & \textbf{0.2524} \\
        \hline
        \textbf{FULLB} & $H_1^{(4)}$ & 0.1083 & 0.0161 & \textbf{0.0030} & 0.0842 \\
        & $H_2^{(4)}$ & 0.4858 & \textbf{0.0038} & 0.0348 & 0.0685 \\
        & $H_3^{(4)}$ & 0.2719 & 0.0102 & \textbf{0.0059} & 0.0684 \\
        & $H_4^{(4)}$ & 0.0065 & \textbf{0.1035} & 0.3928 & 0.1429\\
        & $H_5^{(4)}$ & 0.0901 & \textbf{0.0083} & 0.1379 & 0.0575 \\
        & $H_6^{(4)}$ & 0.0372 & 0.0007 & \textbf{0.0003} & 0.0439 \\
        \hline
        \textbf{SOCFB} & $H_1^{(4)}$ & 0.5283 & 0.1137 & \textbf{0.0051} & 0.3652 \\
        & $H_2^{(4)}$ & 0.4279 & 0.0815 & \textbf{0.0094} & 0.3622 \\
        & $H_3^{(4)}$ & 0.0393 & 0.1187 & \textbf{0.0043} & 0.3287 \\
        & $H_4^{(4)}$ & 0.0018 & \textbf{0.1931} & 0.3014 & 0.4095 \\
        & $H_5^{(4)}$ & 0.0022 & \textbf{0.1172} & 0.2383 & 0.2368 \\
        & $H_6^{(4)}$ & 0.0001 & \textbf{0.0668} & 0.0682 & 0.2652 \\
        \hline
    \end{tabular}
\end{center}
\caption{Graphlet frequencies for all networks with relative error PSRW, Waddle, and Unordered Lifting after 40K graphlet samples (including rejections for Waddle).}
\label{tab:relative_errors}
\vspace{-0.7cm}
\end{figure}

\begin{figure}
    \centering
    \scalebox{.47}{\input{fig1.tex}}
    \scalebox{.47}{\input{fig2.tex}}
    \scalebox{.47}{\input{fig3.tex}}
    \scalebox{.47}{\input{fig4.tex}}
    \scalebox{.47}{\input{fig5.tex}}
    \scalebox{.47}{\input{fig6.tex}}
    \caption{Convergence to the true frequency (shown in black) of the PSRW, Waddle, and Graphlet Lift (GL) methods.}
    \label{fig:converge_to_error}
    \vspace{-0.5cm}
\end{figure}

\begin{figure}
    \centering
    \scalebox{.47}{\input{fig7.tex}}
    \scalebox{.47}{\input{fig11.tex}}
    \scalebox{.47}{\input{fig8.tex}}
    \scalebox{.47}{\input{fig12.tex}}
    \scalebox{.47}{\input{fig9.tex}}
    \scalebox{.47}{\input{fig13.tex}}
    \scalebox{.47}{\input{fig10.tex}}
    \scalebox{.47}{\input{fig14.tex}}
    \caption{Here we compare the TV performance of Unordered Lift with the PSRW method on graphlets $k=3,4,5,6$. 
    The top four figures show the total variation difference between the estimated counts and the ground truth as a function of iterations on all the data sets.
    The bottom four figures show the total variation between successive graphlet count estimates (i.e.\ $TV(\hat N_m(i-1), \hat N_m(i))$) by each method.}
    \label{fig:TV_convergence}
    \vspace{-0.7cm}
\end{figure}

\subsection{Comparisons on graphlets up to $k=6$}
We can compute the total variation distance between a graphlet frequency distribution ($\hat N_m$) and a target distribution ($N_m$) as
\[
TV(\hat N_m, N_m) = \sum_{m} |\hat N_m - N_m|.
\]
We compare the performance of PSRW and the Unordered Lift with this metric as a function of iterations on all the data sets, with $k=3,4,5,6$.
This comparison is demonstrated in Figure~\ref{fig:TV_convergence}.
For $k=5,6$, as the ground truth is unavailable for these data sets (due to the inability of existing methods to handle such large graphlets), we track the convergence of the total variation difference between successive graphlet distribution estimates.

The $k=3,4$ plots show PSRW outperforming Lift on the SOCFB network, while underperforming on the other data sets. 
We suspect this is because PSRW is adapted to sampling the 3-star, the most common graphlets in SOCFB; accordingly, PSRW performs well on the CAIDA set which is dominated by `3-star' graphlets.
This suspicion is confirmed by the advantage lift has on datasets such as FULLB, which concentrates on the `4-path' graphlet instead of the star. 
In this case, PSRW has trouble converging.
The $k=5,6$ plots demonstrate an approximately equivalent convergence rate between the methods.
Both methods get fast initial gains by obtaining a good estimate of the most common graphlets, while the slow convergence that follows depends on sampling the rare graphlets.
Note that PSRW demonstrates the correlation between its samples here by the `plateau' pattern.
(Note that we omitted Waddling from this comparison because in the case of size $k=5,6$ graphlets there was no clear extension of the Waddle protocol.)

\begin{figure}[th!]
\begin{center}

\minipage{\myplotwidth}
  \includegraphics[width=\linewidth]{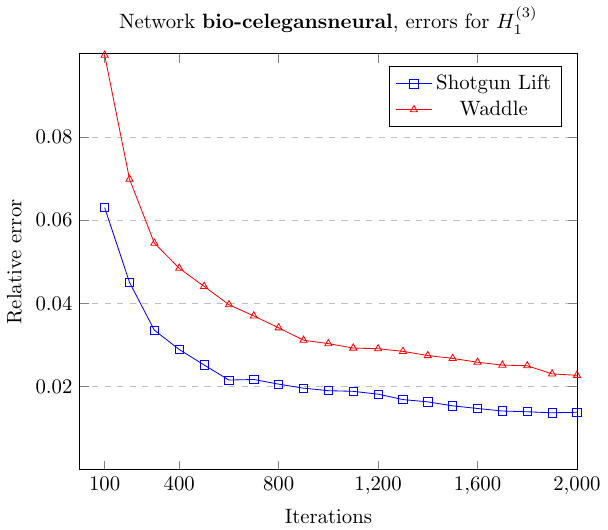}
\endminipage\hfill
\minipage{\myplotwidth}
  \includegraphics[width=\linewidth]{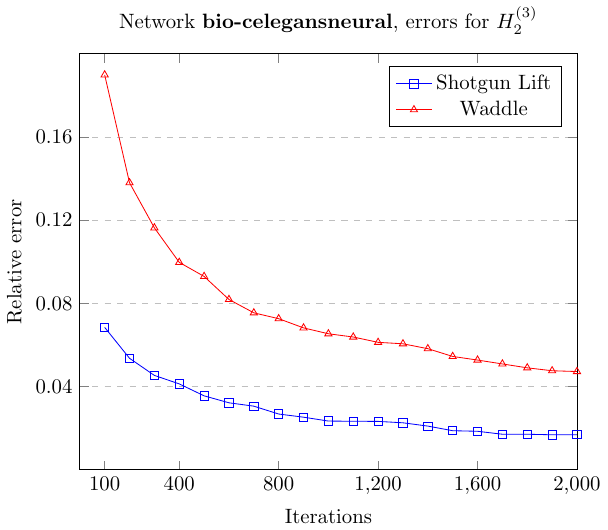}
\endminipage

\hspace*{\fill}
\hspace*{\fill}

\minipage{\myplotwidth}%
  \includegraphics[width=\linewidth]{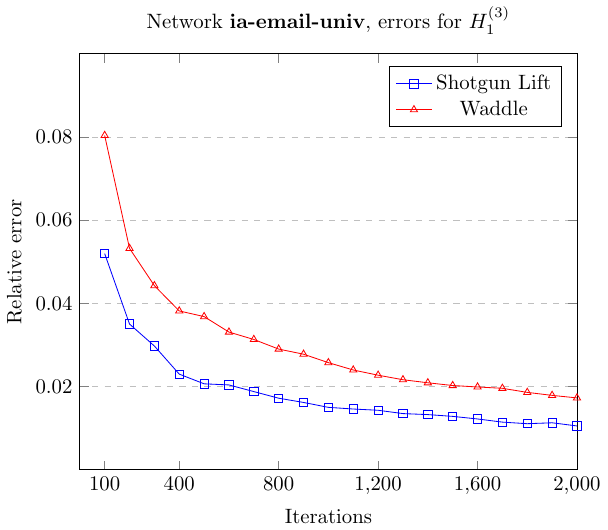}
\endminipage\hfill
\minipage{\myplotwidth}
  \includegraphics[width=\linewidth]{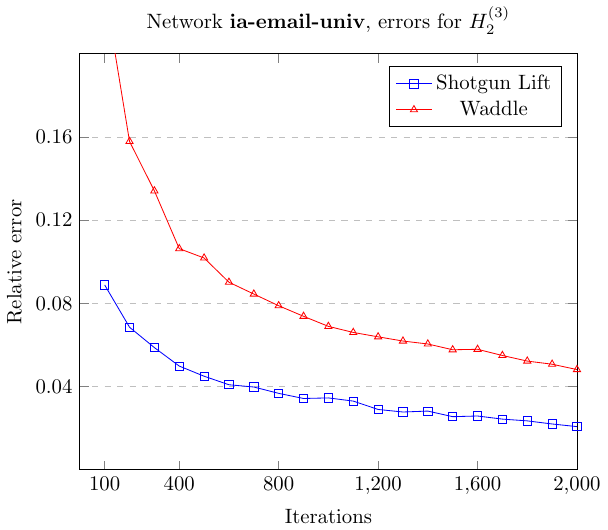}
\endminipage

\hspace*{\fill}
\hspace*{\fill}

\minipage{\myplotwidth}%
  \includegraphics[width=\linewidth]{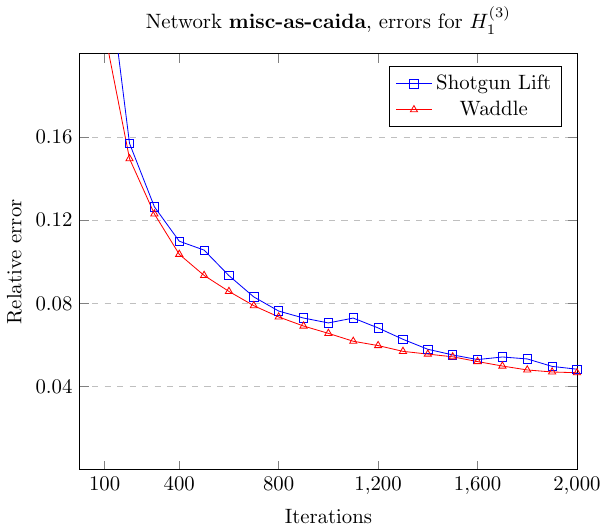}
\endminipage\hfill
\minipage{\myplotwidth}
  \includegraphics[width=\linewidth]{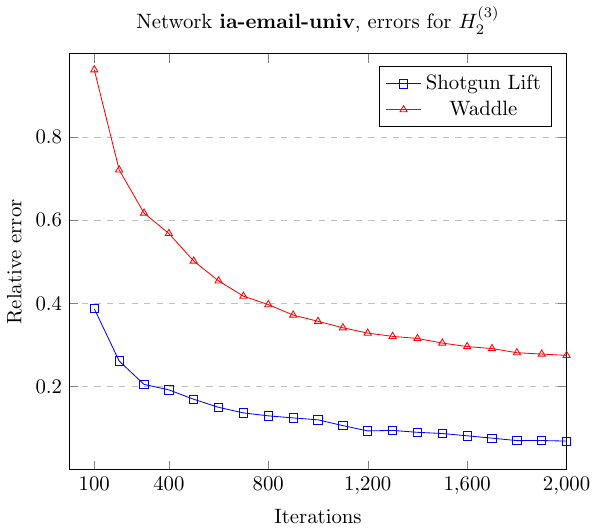}
\endminipage

\hspace*{\fill}
\hspace*{\fill}

\minipage{\myplotwidth}%
  \includegraphics[width=\linewidth]{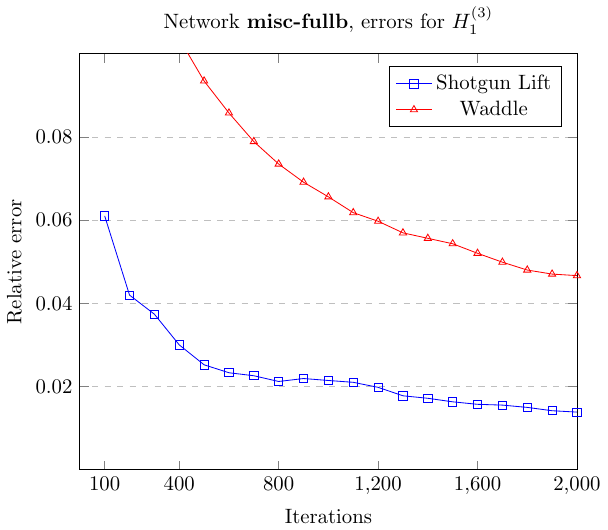}
\endminipage\hfill
\minipage{\myplotwidth}
  \includegraphics[width=\linewidth]{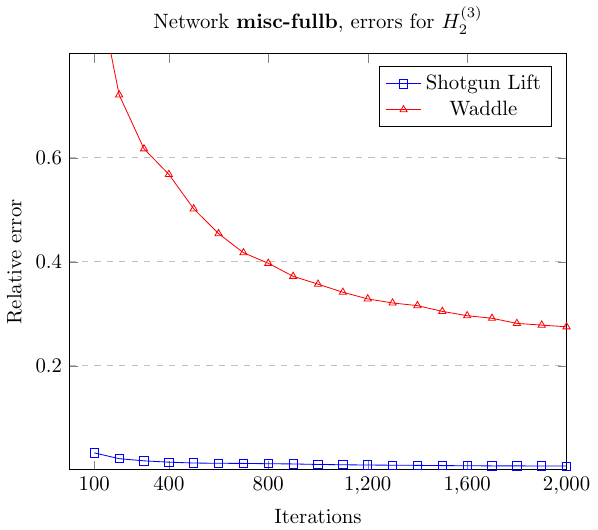}
\endminipage
\hspace*{\fill}
\end{center}
\caption{A comparison of shotgun-Lifting and Waddling for medium sized graphs, $H_1^{(3)}$ (wedge) and $H_2^{(3)}$ (triangle).}
\label{fig:shotgun}
\end{figure}

We also compare the shotgun ordered Lifting relative error against Waddle for the $3$-graphlets, the wedge ($H_1^{(3)}$) and the triangle ($H_2^{(3)}$).
In Figure \ref{fig:shotgun}, we see that the shotgun procedure converges faster than Waddling in these cases.
This advantage comes from shotgun's sampling of many graphlets essentially for free (with the same number of neighborhood queries), we consider all of the graphlets sampled from one shotgun sample to constitute one iteration.
We have observed empirically, that although the shotgun approach produces batches of dependent samples, it is advantageous and we obtain faster convergence.

\section{Conclusion}

A reliable general purpose graphlet sampling algorithm is desireable because it can then be used out of the box without customizations and can scale to massive graphs.
We provide three variants of the Lifting procedure: unordered, ordered, and the shotgun approach.
We showed that the sampling probabilities in Lifting can be calculated from closed form, precomputed functions of the degree sequence of the subgraph.
Lifting exemplifies the characteristics needed for a practical graphlet sampling method: it is easily parallelizable, samples all $k$-graphlets without modification, and can find rare graphlets.
To the best of our knowledge, Lifting is the first graphlet sampling algorithm that enjoys each of these properties.

Our theoretical results bound the variance of Lifting estimated graphlet coefficients, which is based on the largest degrees in the graph.
These results are comparable with the theoretical guarantees for PSRW (after sufficient mixing) and Waddling.
Our experiments demonstrate that Lifting performs well in many cases, obtaining the lowest relative error, particularly for rare graphlets.
We also see that the shotgun procedure can significantly boost the performance without additional neighborhood look-ups.
We conclude by noting that Lifting is able to estimate the $5,6$-graphlet coefficients over a 2.9M vertex graph and the solution converges in total variation in a moderate number of iterations.


\begin{acks}
JS is supported by NSF DMS-1712996.
We are grateful to Peter Dobcs\'anyi for his open-source Pynauty package, which smoothed our Python implementation.
\end{acks}

\bibliographystyle{ACM-Reference-Format}
\bibliography{biblio}

\newpage
\section{Supplement to "Estimating Graphlets via Lifting"}

\subsection{Proof of Prop. \ref{prop:unbiased}.}
    \begin{proof}
    Let $\phi_i$ be as defined in \eqref{phi:ord}, \eqref{phi:shot}.
    For both estimators, because of the form of \eqref{eq:ord_est} and \eqref{eq:shot_est}, if a single term $\phi_i$ is unbiased then $\hat N_m$ is as well.
    Let us begin with $\hat N_{O,m}$, by considering a draw from the lifting process, $A = [v_1,\ldots,v_k]$ which induces the $k$-subgraph, $G|A$.
    By the definition of $\tilde \pi$,
	\begin{multline*}
	    \E \left( \phi_{O,1} \right) = \sum_{A \in V_G^k} \tilde \pi(A) \left( \frac{\ind(T(A) \sim H_m)}{\co(T(A)) \tilde \pi(A)} \right) \\
	    = \sum_{T \in \cV_k} \sum_{A \in V_G^k : T(A) = T} \frac{\ind(T \sim H_m)}{\co(T)} = \sum_{T \in \cV_k} \ind(T \sim H_m) = N_m.  
	\end{multline*}
	Hence, the $\hat N_{O,m}$ is unbiased.
	Consider the shotgun estimator, $\hat N_{S,m}$,
	\begin{multline*}
	    \E \left( \phi_{S,1} \right) = \sum_{B \in V_G^{k-1}} \tilde \pi(B) \sum_{u \in \cN_v(B)} \left( \frac{\ind(G|B \cup \{u\} \sim H_m)}{\co(H_m) \tilde \pi(B)} \right) \\
	    = \sum_{T \in \cV_k} \sum_{B \in V_G^{k-1}} \ind(G|B \cup \{u\} = T, u \in \cN_v(B)) \frac{\ind(T \sim H_m)}{\co(T)} \\
	    = \sum_{T \in \cV_k} \ind(T \sim H_m) = N_m.  
	\end{multline*}
    Hence, the shotgun estimator is unbiased as well.
	\end{proof}

\subsection{Proof of Theorem \ref{thm:var_bd}.}
We can bound the variance in \eqref{eq:variation.ind} by the second moment, which is bounded by,
\begin{equation*}
    \E \phi_1^2 \leq \E \phi_1\max{\phi_1} = N_m(G)\max{\phi_1}.
\end{equation*}
Seeking to control the the maximum of $\phi_1$, we see that,
\begin{equation*}
    \max_T \frac{1}{\pi_U(T)} \le \max_A \frac{1}{|\co(T)| \tilde \pi(A)} \le 
    \max\frac{\prod_{r=1}^{k-1} (d_1+\ldots+d_r)}{|\co(H_m)|\pi_1(d_1)},
\end{equation*}
\begin{equation*}
    \max_B \frac{|\cN_v(B)|}{|\co(H_m)|\tilde\pi(B)} \le 
    \max\frac{\prod_{r=1}^{k-1} (d_1+\ldots+d_r)}{|\co(H_m)|\pi_1(d_1)}.
\end{equation*}
Thus, we can construct a bound on $V_m^{\independent}(\phi_1)$.

\subsection{Mixing time of lifted MCMC}

Let us focus on the sampling vertices via random walk in this subsection.
One advantage of the lifting procedure over the SRW is that it inherits the mixing properties from the vertex random walk.
This can be thought of as a consequence of the data processing inequality in that the lifted CISs are no more dependent then the starting vertices from which they were lifted.
To that end, let us review some basics about mixing of Markov chains,

\begin{definition}
  Define the mixing coefficient of a stationary Markov chain with discrete state space $X_t \in \mathcal X$ as
  \begin{equation}
  \label{eq:gamma}
      \gamma_X(h) = \frac{1}{2}\max_{x_1\in \mathcal X} \sum_{x_2 \in \mathcal X} |\P(X_{t+h} = x_2, X_t = x_1) - \pi(x_1) \pi(x_2)|,
  \end{equation}
  where $\pi(x)$ is the stationary distribution of the Markov chain.
  Also, define the mixing time of a stationary Markov chain $\{X_t\}$ as
  \begin{equation}
      \tau_X(\varepsilon) = \min\left\{\, h \mid \gamma_X(h) < \varepsilon  \right\}.
  \end{equation}
\end{definition}

\begin{theorem}\cite{Sinclair1992}
  Given stationary Markov chain $\{X_t\}$ with $\mu < 1$ being the second largest eigenvalue of the transitional matrix,
  \begin{equation}
      \gamma_X(h) \leq e^{-(1-\mu)h}.
  \end{equation}
\end{theorem}

There are two consequences of mixing for CIS sampling.
First, an initial burn-in period is needed for the distribution $\pi$ to converge to the stationary distribution (and for the graphlet counts to be unbiased).
Second, by spacing out the samples with intermediate burn-in periods and only obtaining CISs every $h$ steps we can reduce the covariance component of the variance of $\hat N_m$.
Critically, if we wish to wait for $h$ steps, we do not need to perform the lifting scheme in the intervening iterations, since those graphlets will not be counted.
So, unlike in other MCMC method, spacing in lifted CIS sampling is computationally very inexpensive.
Because burn-in is a one-time cost and requires only a random walk on the graph, we will suppose that we begin sampling from the stationary distribution, and the remaining source of variation is due to insufficient spacing between samples.
The following theorem illustrates the point that the lifted MCMC inherits mixing properties from the vertex random walk.

\begin{theorem}
\label{thm:cor_bd}
Consider sampling a starting vertex from a random walk, such that a sufficient burn in period has elapsed and stationarity has been reached.
Let $h$ be the spacing between the CIS samples, $D$ be defined as in Theorem \ref{thm:var_bd}, and $\mu$ be the second largest eigenvalue of the transition matrix for the vertex random walk.
Let $\phi_i$ be as defined in \eqref{phi:ord}, \eqref{phi:shot} or \eqref{phi:unord}, then
\begin{equation*}
    \left\vert \Cov\left(\phi_i, \phi_{i+1})\right)\right\vert \leq
    8 N_m(G)|E_G|^2 e^{- (1-\mu) h} D.
\end{equation*}
\end{theorem}

\begin{corollary}
In the notation of the Theorem~\ref{thm:cor_bd},
\begin{equation*}
    \frac{2}{n} \left\vert\sum_{i < j} \Cov\left(\phi_i, \phi_j\right) \right\vert \leq
     8 N_m(G)|E_G|^2 \frac{e^{- (1-\mu) h}}{1- e^{- (1-\mu) h}} D.
\end{equation*}
\end{corollary}

Hence, if we allow $h$ to grow large enough then we can reduce the effect of the covariance term, and our CISs will seem as if they are independent samples.

\begin{figure}[th!]
\begin{center}

\minipage{\myplotwidth}
  \includegraphics[width=\linewidth]{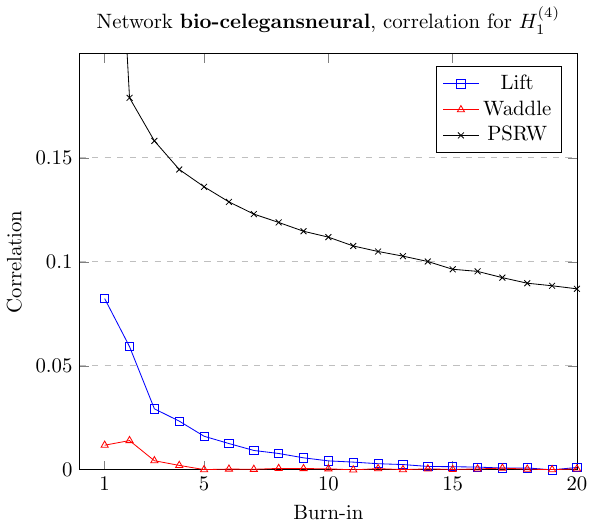}
\endminipage\hfill
\minipage{\myplotwidth}
  \includegraphics[width=\linewidth]{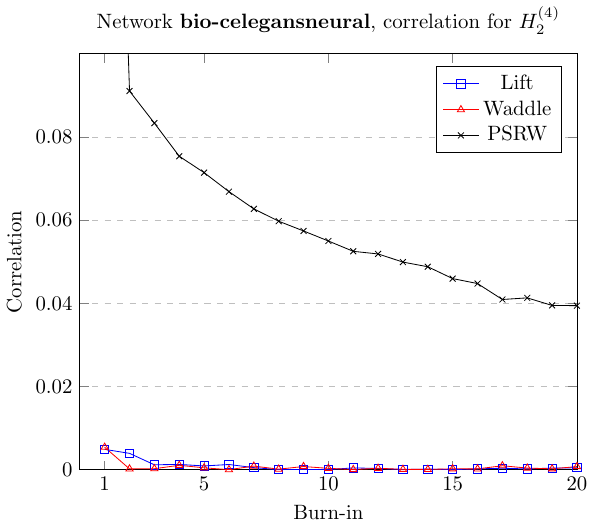}
\endminipage

\hspace*{\fill}
\vskip 1pt
\hspace*{\fill}

\minipage{\myplotwidth}%
  \includegraphics[width=\linewidth]{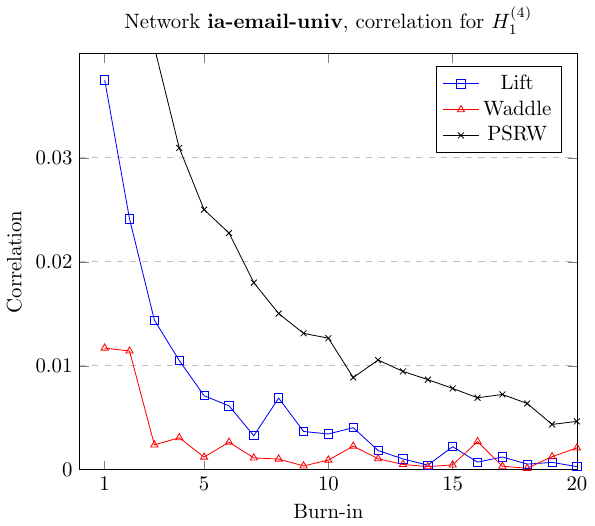}
\endminipage\hfill
\minipage{\myplotwidth}
  \includegraphics[width=\linewidth]{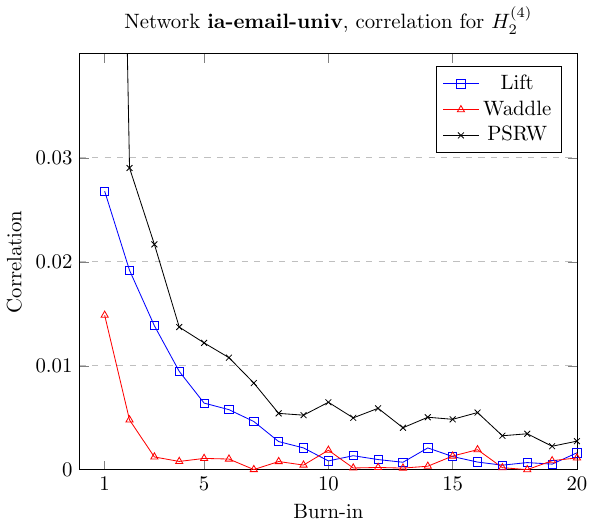}
\endminipage

\hspace*{\fill}
\vskip 1pt
\hspace*{\fill}

\minipage{\myplotwidth}%
  \includegraphics[width=\linewidth]{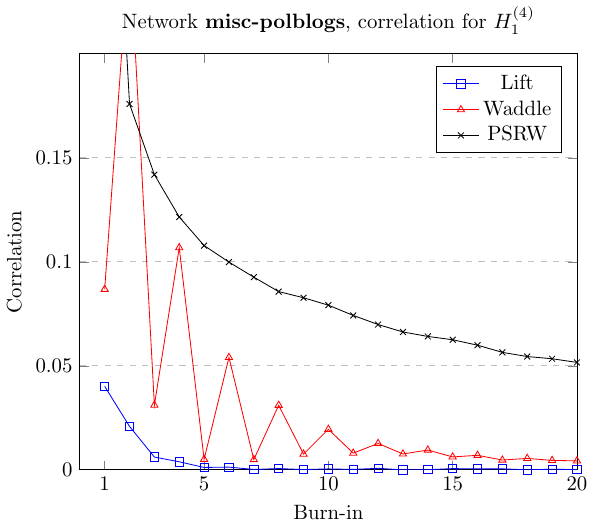}
\endminipage\hfill
\minipage{\myplotwidth}
  \includegraphics[width=\linewidth]{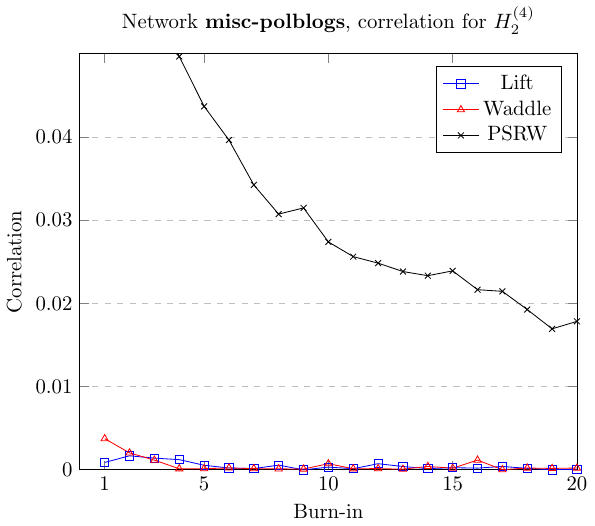}
\endminipage

\hspace*{\fill}
\vskip 1pt
\hspace*{\fill}

\minipage{\myplotwidth}%
  \includegraphics[width=\linewidth]{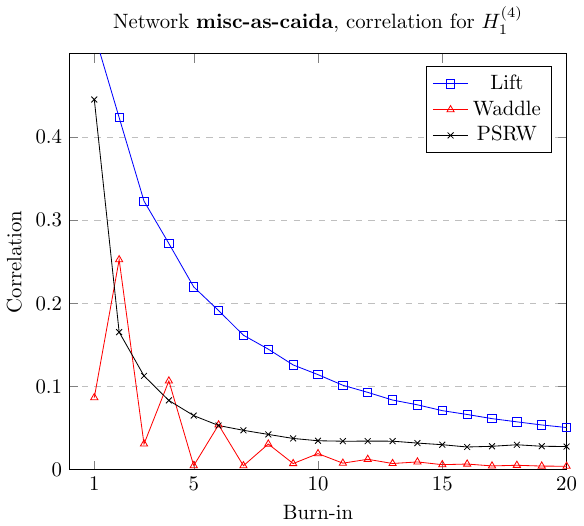}
\endminipage\hfill
\minipage{\myplotwidth}
  \includegraphics[width=\linewidth]{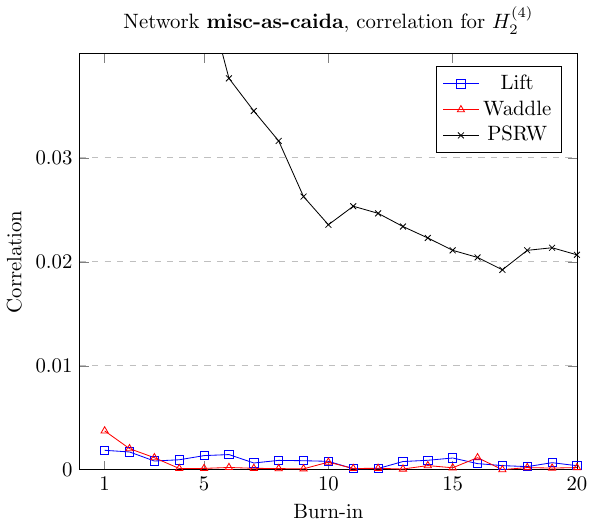}
\endminipage

\hspace*{\fill}
\vskip 1pt
\hspace*{\fill}

\minipage{\myplotwidth}%
  \includegraphics[width=\linewidth]{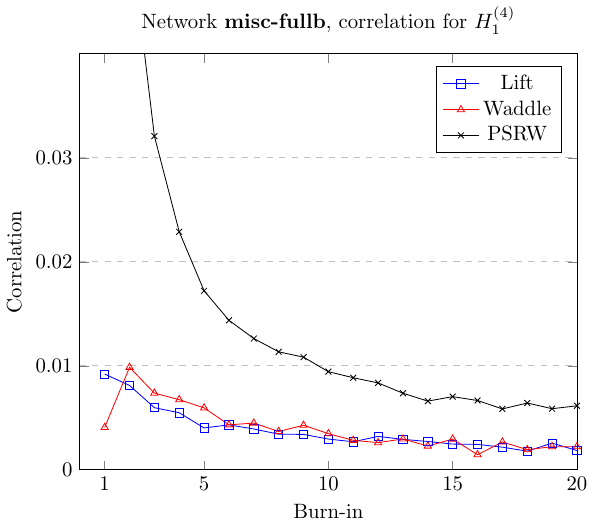}
\endminipage\hfill
\minipage{\myplotwidth}
  \includegraphics[width=\linewidth]{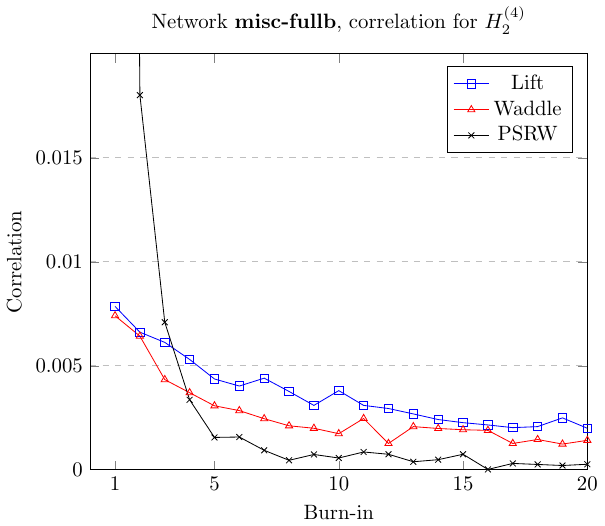}
\endminipage

\hspace*{\fill}
\end{center}
\caption{Correlation of $\phi_i$ and $\phi_{i+1}$ depending on the intermediate burn-in time, $h$, between samples for graphlets $H_1^{(4)}$ and $H_2^{(4)}$.}
\label{fig:correlation}
\end{figure}

Next, for the random walk lifting, we empirically compare the dependence of $\phi_i$ and $\phi_{i+1}$ using correlation for different values of the burn-in $h$ (see Fig.\ref{fig:correlation}).
For Lift and Waddling, the burn-in between $\phi_i$ and $\phi_{i+1}$ is the number of steps taken after sampling $T_i$ to get a new starting vertex for $T_{i+1}$.
For PSRW, burn-in is the number of steps between CIS samples in the random walk on subgraphs.
From the graphs in Figure~\ref{fig:correlation}, we see that PSRW produces highly correlated samples compared to Lift and Waddling methods.
This agrees with our analysis of PSRW, since it takes many more steps for the subgraph random walk to achieve desired mixing compared to the random walk on vertices.

\subsection{Proof of Theorem \ref{thm:cor_bd}}

Let $\phi_i$ be as defined in \eqref{phi:ord}, \eqref{phi:shot} or \eqref{phi:unord}.
Given two starting vertices $v_i$ and $v_j$ of the lifting process, notice that random variables $\phi_i|v_i$ and $\phi_j|v_j$ are independent.
Therefore
\begin{multline*}
    \E\left(\phi_i \phi_{i+1})\right)=
    \E_{\pi_1(v_i)\times \pi_1(v_{i+1})} \E\left(\phi_i \phi_{i+1}| v_i, v_{i+1}\right) =\\
     \E_{\pi_1(v_i)\times \pi_1(v_{i+1})} \left(\E\left(\phi_i|v_i\right) \E\left(\phi_{i+1}|v_{i+1}\right)\right).
\end{multline*}

Using the equation above, we can bound the covariance of $\phi_i$ and $\phi_{i+1}$ with basic inequalities:
\begin{align*}
    |\Cov\left(\phi_i, \phi_{i+1}\right)| \leq & \\
    \sum_{x_1,x_2\in V_G}  \E(\phi_i|v_i = & x_1)  \E\left(\phi_{i+1}|v_{i+1}=x_2\right)\\
    &\left\vert \P(v_i=x_1,v_{i+1}=x_2) - \pi_1(x_1)\pi_1(x_2)\right\vert\leq \\
    \max_{x_2\in V_G}  \E(\phi_{i+1}|v_{i+1} & =x_2) \sum_{x_1} \E\left(\phi_i|v_i = x_1\right)\\
    \max_{x_1} & \sum_{x_2}  \left\vert \P(v_i=x_1,v_{i+1}=x_2) - \pi(x_1)\pi(x_2)\right\vert = \\
      2\gamma_{G_V}(h) \max_{x_2} &\ \E\left(\phi_{i+1}|v_{i+1}=x_2\right) \sum_{x_1} \E\left(\phi_i|v_i = x_1\right),
\end{align*}

where $\gamma_{G_V}(h)$ is the mixing coefficient from \eqref{eq:gamma} for the random walk on vertices.
Next, estimate factors from the RHS as follows:
\begin{multline}
    \sum_{x}\E\left(\phi|v = x\right) \leq
    \max_{x}\frac{1}{\pi(x)} \sum_{x}\E\left(\phi|v = x\right) \pi(x)\leq \\
    2|E_G| N_m(G).
\end{multline}

For $\max_x \E\left(\phi|v=x\right)$, consider the expressions for $\phi$ from \eqref{phi:ord}, \eqref{phi:shot} or \eqref{phi:unord}.

Using notation $D = \prod_{r=2}^{k-1} (\Delta_1 +\ldots + \Delta_r)$, for the Ordered Lift estimator,
\begin{multline*}
    \max_x \E\left(\phi_O|v=x\right) \le \max_x \sum_{A} \frac{\P(A|v=x)}{\tilde\pi(A)} \le \\
    \max_x \frac{|\{A\mid A[1]=x\}|}{\pi(x)} \le 2|E_G|D.
\end{multline*}
For the Shotgun Lift estimator,
\begin{multline*}
    \max_x \E\left(\phi_S|v=x\right) \le
    \max_x \sum_B |\cN_v(B)|\frac{\P(B|v=x)}{\tilde\pi(B)} \leq \\
    \max_x \frac{|\cN_v(B)| |\{B\mid B[1]=x\}|}{\pi(x)} \leq
    2|E_G| D,
\end{multline*}
For the Unordered Lift estimator,
\begin{multline*}
    \max_x \E\left(\phi_U|v=x\right) \le
    \max_x \sum_T \frac{\P(T|v=x)}{\pi_U(T)} \leq \\
    \max_x \frac{|\{T\mid x\in V_T\}|}{\pi(x)} \leq
    2|E_G| D.
\end{multline*}

Combining the results, we get the desired bound.

\end{document}